# P. W. Bridgman contributions to the foundations of shock compression of condensed matter

W J Nellis

Department of Physics, Harvard University, Cambridge MA 02138, USA

E-mail: nellis@physics.harvard.edu

Abstract. Based on his 50-year career in static high-pressure research, P. W. Bridgman (PWB) is the father of modern high-pressure physics. What is not generally recognized is that Bridgman was also intimately connected with establishing shock compression as a scientific tool and he predicted major events in shock research that occurred up to 40 years after his death. In 1956 the first phase transition under shock compression was reported in Fe at 13 GPa (130 kbar). PWB said a phase transition could not occur in a ~microsec, thus setting off a controversy. The scientific legitimacy of shock compression resulted 5 years later when static high-pressure researchers confirmed with x-ray diffraction the existence of epsilon-Fe. Once PWB accepted the fact that shock waves generated with chemical explosives were a valid scientific tool, he immediately realized that substantially higher pressures would be achieved with nuclear explosives. He included his ideas for achieving higher pressures in articles published a few years after his death. L. V. Altshuler eventually read Bridgman articles and pursued the idea of using nuclear explosives to generate super high pressures, which subsequently morphed today into giant lasers. PWB also anticipated combining static and shock methods, which today is done with pre-compression of a soft sample in a diamond anvil cell followed by laser-driven shock compression. One variation of that method is the reverberating-shock technique, in which the first shock pre-compresses a soft sample and subsequent reverberations isentropically compress the first-shocked state.

1. Introduction
P. W. Bridgman published over 200 papers between 1906 and 1961 on the properties of materials at static pressures as high as 20 GPa. Bridgman was Professor of Physics at Harvard. In 1946 he won the Nobel Prize in Physics. His most important papers have been collected in seven volumes [1]. His graduate students and (year graduated) included John C. Slater (1929), John Van Vleck (1923), Francis Birch (1932), and Gerald Holton (1948). John Slater was a professor at MIT and is well known for the physics textbooks he wrote and his contributions to Quantum Mechanics. John Van Vleck and Francis Birch were professors at Harvard. Van Vleck was awarded the 1977 Nobel Prize in Physics for his research on magnetism, a Prize he shared with Philip Anderson and Sir Neville Mott. Birch derived the first equation of state for materials at 100 GPa pressures in the deep Earth. Holton is Professor of Physics and of the History of Science, Emeritus, at Harvard.

During the Second World War Bridgman did high-pressure experiments as part of the Manhattan Project. After the War, Los Alamos began performing Hugoniot experiments on many materials up to several tens of GPa using chemical explosives. A Hugoniot experiment is one in which the velocity of a shock front $u_s$ generated in a material is measured and the velocity of material $u_p$ just behind the shock front is deduced from measured quantities. Conservation of momentum, mass, and internal energy across the shock front and the assumption that the compression is adiabatic allows calculation of shock pressure P, density ρ, and internal energy E. The detectors were electrical pins, which measured arrival times of a shock front at various points in a sample. Experimental lifetimes were ~microsec and detector resolution was ~0.01 microsec.

2. Verification of the shock-induced alpha-epsilon transition in Fe
In 1956 Bancroft et al measured the Hugoniot of Fe up to 20 GPa using ~65 pins on each of several shots [2]. A large number of pins were used because two and possibly three shock waves were expected, one caused by elastic strength, one caused by plastic compression above the Hugoniot elastic limit, and perhaps a third wave if a phase transition were to occur, all three having successively slower shock velocities. They discovered three distinct waves and attributed the third wave to a phase transition at 13 GPa from alpha-bcc Fe at ambient to a then unknown crystal structure. They estimated the temperature at the transition to be 40 C or less, which is essentially room temperature. Hence, in principal the transition reported by Bancroft et al could be checked by measurements at static high pressures.

Bridgman, the leading high-pressure researcher in the world at that time, said it was highly unlikely that a transition with a change in crystal structure could occur "in times as short as a few microseconds". He tried to verify the existence of the phase transition by measuring electrical resistance of Fe up to 17 GPa. He found no indication of a phase transition in electrical resistance measurements at any pressure. While acknowledging the resistance method was not definitive, he said the observation of a third shock wave probably needed to be explained by something other than a phase transition [3].

In 1962 John Jamieson and A. W. Lawson reported development of a high-pressure cell in which x-ray diffraction patterns could be measured in specimens under quasi-hydrostatic pressures up to 15 GPa or higher. At their highest pressures they observed that alpha-Fe transforms to a new phase, which they identified as hcp, now known as epsilon-hcp [4]. At about the same time Harry Drickamer and A. L. Balchan measured the electrical resistance of Jamieson's Fe sample and detected a phase transition at 13 GPa [5], the same pressure of the phase transition reported by Bancroft et al. The resolution of the question of the alpha-epsilon transition in Fe established shock compression as a valid scientific tool.

Bridgman questioning of the interpretation of the shock-wave data of Bancroft et al had set off a scientific controversy that had taken five years to resolve. His high standing in the high-pressure community motivated people in static-pressure research to expand substantially their experimental capabilities with x-ray diffraction. In so doing shock and static researchers have worked on issues of common importance from the earliest days of shock research.

3. High pressure awards established
In 1977 AIRAPT, the International Association for the Advancement of High Pressure Science and Technology, established the P. W. Bridgman Award for life-time achievement. Harry Drickamer was its first recipient. Francis Birch received the Bridgman Award in 1981. In the mid 1980s the John Jamieson Award was established for significant achievement by a graduate student or postdoctoral researcher. The trust that supports the Jamieson Award was established and is managed by Bob Schock, a geophysicist at the Lawrence Livermore National Laboratory and a former postdoc of Jamieson. The Jamieson Award is presented by both by the High Pressure Gordon Conference in the U.S. and by AIRAPT.

In 1987 the Topical Group on Shock Compression of Condensed Matter of the American Physical Society established the G. E. Duvall Award for life-time achievement. On the plaque that accompanies this award is a plot of the pressure-volume Hugoniot curve of Fe showing the alpha-epsilon transition to commemorate its importance in high-pressure research and, implicitly, Bridgman initiation of the controversy that led to its understanding.

4. Bridgman predictions about shock-compression research

Once Bridgman realized that chemical explosives could be used reliably to achieve high pressures and observe phase transitions, he shortly thereafter realized that nuclear explosives could be used to achieve even higher pressures. On the last page of the seventh and last volume of his collected work he wrote the following: "The very highest pressures will doubtless continue to be reached by some sort of shock-wave technique…Perhaps some fortunate experimenters may ultimately be able to command the use of atomic explosives in studying this field [6,7]." Bridgman died in August of 1961. Bridgman's suggestion was eventually read and pursued by L. V. Altshuler and his group at Arzamas-16. In his memoirs Altshuler writes [8]: In 1962 Bridgman…suggested that with some luck, experimenters might even employ atomic blasts in high-pressure research. Such lucky experimenters were my team's members…who in 1968 were the first to carry out measurements in the near zone of an underground nuclear explosion [9]. Today, with the Comprehensive Test Ban Treaty in place, shock experiments at ultrahigh pressures are done with pulsed giant lasers.

In the same paper Bridgman states: It is conceivable that a way will be found of superimposing shock-wave pressures with static pressures [6,7]. This kind of experiment is done currently by compressing a sample statically in a diamond anvil cell (DAC) and then shock compressing it to high pressure generated with a pulsed giant laser. For example, water has been pre-compressed to 1 GPa in a DAC and then shock-compressed with a laser to 250 GPa [10]. A variation of the above pre-compression experiment is a reverberating shock-wave experiment in which the first shock pre-compresses a liquid to a relatively low shock pressure and the remaining reverberating shocks then compress the fluid isentropically from that first-shock state. Essentially the same states are achieved in water up to 100 GPa using both methods [10,11]. The reverberating shock method was also used to produce metallic fluid hydrogen [12].

To complete the Fe story, in 2005 an x-ray beam was split into two beams, one to generate an intense x-ray source and one to generate a shock in a small, oriented single crystal of Fe. The experimental results showed that a phase transition occurred at the expected pressure and the structure of the high- pressure phase is in fact hcp [**13**], the same phase observed by Jamieson and Lawson in 1961.

5. Conclusions

Bridgman was truly remarkable in terms of his contributions to static high-pressure research over his 50-year career, the students he produced, his important contributions to establishing the scientific foundations of shock-compression research, and his predictions of advances in shock-compression research that occurred up to forty years after his death.

6. Acknowledgment

I wish to acknowledge B. Altshuler, the son of L. V. Altshuler, for useful discussions about the history of using nuclear explosives to achieve ultrahigh shock pressures.